\documentclass[conference,compsoc]{IEEEtran} %\documentclass[conference]{IEEEtran}
\usepackage[utf8]{inputenc}
\usepackage{cite}
\usepackage{amsmath,amssymb,amsfonts}
\usepackage{algorithmic}
\usepackage{graphicx}
\usepackage{textcomp}
\usepackage{xcolor}

\usepackage{url}
\usepackage{booktabs}
\usepackage{algorithm}
\usepackage{listings}
\usepackage{multirow}
\newcommand{\system}{BARS}

\def\BibTeX{{\rm B\kern-.05em{\sc i\kern-.025em b}\kern-.08em
    T\kern-.1667em\lower.7ex\hbox{E}\kern-.125emX}}
\begin{document}

\title{BARS: Benign-Anchored Ranking and Selection for False Alarm Reduction in Network Intrusion Detection\\
}

\author{
% \IEEEauthorblockN{1\textsuperscript{st} Anonymous}
% \IEEEauthorblockA{\textit{Anonymous} \\
% }
\IEEEauthorblockN{1\textsuperscript{st} Abu Fuad Ahmad}
\IEEEauthorblockA{\textit{Department of Computer Science} \\
\textit{New Mexico State University}\\
Las Cruces, New Mexico, USA\\
fuad@nmsu.edu}
\and
\IEEEauthorblockN{2\textsuperscript{nd} Istiaque Ahmed}
\IEEEauthorblockA{\textit{Graduate School of Informatics} \\
\textit{Osaka Metropolitan University}\\
Osaka, Japan\\
sw23837u@st.omu.ac.jp}
}

\maketitle

%----------------------------------------------Opss
\begin{abstract}
False alarms remain one of the primary barriers to the operational
deployment of network intrusion detection systems (NIDS): in
high-volume environments, even a sub-1\% false positive rate produces
tens of thousands of daily alerts, overwhelming security analysts and
eroding trust in automated detection. Filter-based feature selection
is an attractive lever for false alarm reduction because it operates
upstream of the classifier and incurs no inference-time cost. Yet
classical filters apply class-symmetric criteria that ignore the
operational asymmetry of intrusion detection, in which benign traffic
defines the baseline and attacks are deviations from it. A recent
class-asymmetric filter, Classwise Mean Deviation (CMD), addresses
this asymmetry but anchors its score to a global reference mean that
drifts toward attack distributions under imbalance, attenuating
precisely the deviations it is meant to capture.

We propose \emph{Benign-Anchored Ranking and Selection} (BARS), a
two-stage filter that replaces CMD's global anchor with the
benign-class mean and follows the score with an order-preserving
decorrelation step. We evaluate BARS on three NIDS benchmarks spanning
the imbalance spectrum (CICIDS2017, CICDDoS2019, UNSW-NB15) across
five feature budgets $k \in \{5, 10, 20, 30, 40\}$. The empirical
pattern matches the design rationale: on attack-majority data, where
the global-anchor bias is most severe, BARS reduces FPR over CMD by
15.4\% on UNSW-NB15 at $k=20$ and by 21--23\% on CICDDoS2019 at small
budgets, while preserving true positive rate and macro-F1. On
benign-majority data, BARS and CMD converge, consistent with the
theoretical limit in which the global and benign-anchored scores
coincide. \system{} reduces false alarms by up to 32\% while maintaining similar detection rates, with larger gains under stronger imbalance.
Ablation results show that the two stages provide complementary benefits and
select a consistent set of robust features. BARS's linear-time scoring and
low memory footprint make it suited to deployments where richer
methods cannot run.
\end{abstract}


\section{Introduction}
\label{sec:introduction}
The rapid growth of modern communication networks, cloud services, IoT devices, and digital applications has generated massive volumes of network traffic data, making network security analytics a prominent big data challenge. Such datasets are characterized not only by their scale but also by high dimensionality, feature redundancy, class imbalance, and the frequent presence of missing values caused by packet loss, sensor failures, or incomplete data collection \cite{buczak2015surveyMLcyber,ahmad2021review,sayeed2024smartic}. These issues can significantly degrade the performance of machine learning models by increasing computational complexity, prolonging training time, reducing model interpretability, and introducing uncertainty into the learning process \cite{ SommerPaxson2010,ahmad2024survayMVI,ahmad2024impactMVI}. Furthermore, noisy, redundant, and incomplete features often contribute to higher false alarm rates, a critical concern in intrusion detection systems where excessive false positives can overwhelm security analysts and reduce trust in automated defenses. 
%Network intrusion detection systems (NIDS) are a core component of modern cybersecurity infrastructure, tasked with identifying malicious activity in high-volume network traffic. 
Although machine-learning-based Network intrusion detection systems (NIDS) have substantially improved aggregate detection metrics over the
past decade, their operational adoption remains constrained by a more
fundamental problem: \emph{false alarms}. In production deployments,
the dominant bottleneck is rarely whether attacks are detected, but
the rate at which benign traffic generates alerts. A NIDS operating at
a 1\% false positive rate (FPR) on a network handling $10^7$ benign
flows per day produces on the order of $10^5$ false alerts daily---well
beyond the triage capacity of typical security operations
teams~\cite{SommerPaxson2010}. The downstream
consequences---analyst fatigue, alert suppression, slowed incident
response, and eroded trust in automated detection---are well
documented~\cite{BuczakGuven2016Survey}.

Two intrinsic properties of network traffic data amplify the severity
of false alarms. First, NIDS datasets are typically
\emph{high-dimensional}, with dozens to hundreds of flow-level features
extracted from each connection~\cite{Ring2019Survey}. Second, they are
\emph{often imbalanced}, with class proportions varying widely across
benchmarks and deployments~\cite{He2009Imbalanced}. Under these
conditions, learning-based models tend to overfit to subtle variations
within benign traffic, producing spurious detections and elevated FPR.

\textbf{Why feature selection.}
Existing approaches to false alarm reduction operate predominantly at
the classifier or decision stage---through cost-sensitive learning,
threshold tuning, or anomaly-score
calibration~\cite{Fawcett2006ROC,Liu2008IsolationForest,Scholkopf2001OCSVM}.
These approaches share a structural limitation: they compensate for
noise post-hoc rather than eliminating it at the source. Feature
selection, in contrast, operates upstream of the classifier. By
removing features that vary within benign traffic but carry no
attack-discriminative signal, feature selection reduces the noise
floor that classifiers must contend with, without introducing
additional inference-time cost or coupling to any particular learning
algorithm.

\textbf{Why existing filters fall short.}
Classical filter methods such as Mutual
Information~\cite{vergara2014mi}, Pearson
correlation~\cite{nasir2020pearson}, and Fisher
Score~\cite{gu2012fisher} rank features using globally symmetric
statistics that treat all classes equivalently. This symmetry is
misaligned with the structure of intrusion detection, where benign
traffic defines the operational baseline and attacks are naturally
characterized as \emph{deviations} from that baseline. Recent work
introduced Classwise Mean Deviation
(CMD)~\cite{AhmadICMLA2025}, which uses class-wise mean shifts for
feature ranking and achieves competitive performance in NIDS tasks.
However, CMD anchors its scoring to a global reference mean. Under
class imbalance, this global mean drifts toward the dominant class,
producing two related effects: when benign dominates, a non-zero
benign self-deviation that rewards features varying within benign
traffic; when attacks dominate, systematically attenuated attack-class
deviations that underweight precisely the features CMD is meant to
capture.

\textbf{Our approach.}
The core idea behind our method is to anchor feature scoring directly
to the benign class rather than to a global reference. A benign anchor
coincides with the operational baseline a NIDS must respect,
eliminates the self-deviation that global-mean anchoring assigns to
benign-varying features, and preserves the full magnitude of
attack-class deviations regardless of training-time class composition.
We embed this principle in \emph{Benign-Anchored Ranking and
Selection} (BARS), a two-stage filter: Stage~1 ranks features by their
absolute mean deviation from benign traffic; Stage~2 walks the ranked
list and admits each feature only if its correlation with all
previously admitted features stays below a single threshold $\tau$.
The resulting method is classifier-agnostic, governed by one
interpretable hyperparameter, and computes its scoring in
$\mathcal{O}(Nd)$ time. We evaluate BARS on three NIDS benchmarks
spanning the imbalance spectrum (CICIDS2017, CICDDoS2019, UNSW-NB15)
and find that the empirical pattern matches the design rationale: on
attack-majority data, where the global-anchor bias is most severe,
BARS reduces FPR over CMD by $15$--$23\%$ depending on dataset and
budget; on benign-majority data, BARS and CMD converge, consistent
with the theoretical limit in which the two scores coincide.

\textbf{Contributions.}
This paper makes the following contributions:
\begin{itemize}
    \item We identify a structural failure mode in Classwise Mean
    Deviation (CMD), a recently proposed filter method for NIDS: its
    global anchor drifts toward the dominant class under imbalance,
    attenuating the very class-deviation signal it is designed to
    capture.

    \item We propose BARS, which replaces CMD's global anchor with
    the benign-class mean and adds an order-preserving decorrelation
    step to prevent redundant feature selection. BARS is
    parameter-light, classifier-agnostic, and computes its scoring in
    $\mathcal{O}(Nd)$ time.

    \item We show empirically that BARS's advantage materializes
    precisely where the design rationale predicts: on attack-majority
    data, BARS reduces FPR over CMD by 15--23\% with detection
    capability preserved; on benign-majority data, the two methods
    converge.
\end{itemize}

\textbf{Scope.}
We position BARS narrowly. It is a principled refinement of CMD
specifically, not a universally dominant filter: classical methods
with richer scoring functions (Pearson, Mutual Information) achieve
lower FPR than BARS on several (dataset, budget) combinations.
BARS's distinguishing value lies in its linear-time scoring and
low memory footprint---properties most useful in deployments where
richer methods cannot run, as evidenced by Fisher Score and Mutual
Information exceeding $1$\,TB of memory on the larger benchmarks in
our evaluation.
    
The remainder of this paper is organized as follows.
Section~\ref{sec:background} reviews the NIDS pipeline, formalizes our
threat model, and recaps CMD. Section~\ref{sec:related} surveys
related work in feature selection, false alarm reduction, and
imbalanced learning. Section~\ref{sec:methodology} introduces BARS
and its design rationale. Section~\ref{sec:experimental_setup}
describes the experimental setup, and Section~\ref{sec:results}
presents the empirical results. Section~\ref{sec:discussion} discusses
when benign anchoring helps, addresses adaptive adversaries, and
considers deployment, and Section~\ref{sec:conclusion} concludes.

\section{Background and Threat Model}
\label{sec:background}

This section describes the NIDS pipeline in which BARS operates,
formalizes the threat model, and recaps the Classwise Mean Deviation
(CMD) method that BARS extends.

\subsection{Network Intrusion Detection Pipeline}
\label{subsec:nids_pipeline}

A modern NIDS processes network traffic in four sequential stages:
\begin{enumerate}
    \item \emph{Traffic capture}---raw packets or flows are collected
    from network sensors.
    \item \emph{Feature extraction}---statistical and protocol-level
    features are computed over flow windows (e.g., packet counts,
    inter-arrival times, byte distributions).
    \item \emph{Feature selection or reduction}---a subset of
    informative features is retained to reduce computational cost and
    overfitting.
    \item \emph{Classification}---a learned model assigns each flow to
    a benign or attack class, and alerts are forwarded to a Security
    Operations Center (SOC) for analyst triage.
\end{enumerate}

BARS operates at stage~(iii). It is a univariate filter that scores
features prior to classifier training and is agnostic to the
downstream learning algorithm, making it a drop-in replacement for
existing filter methods such as Mutual Information or Fisher Score.
Because the operational cost of false alarms is realized at
stage~(iv), any reduction achieved at stage~(iii) propagates directly
to operational savings without coupling to classifier-specific design
choices.

\subsection{Threat Model}
\label{subsec:threat_model}

We consider a defender who operates a NIDS trained on a labeled
corpus of historical benign and attack traffic and whose goal is to
reduce the operational alert burden---i.e., minimize the false
positive rate (FPR) while maintaining the true positive rate (TPR)
above an acceptable threshold. The defender has full access to the
training pipeline (feature extraction, feature selection, classifier
training) and selects BARS as the feature selection component.

The adversary seeks to inject malicious traffic---denial-of-service,
probing, exploitation, or data-exfiltration flows---without being
flagged, and is assumed to follow attack patterns drawn from known
families and commodity tooling. The adversary has no access to the
training data, learned parameters, or selected feature subset of the
deployed NIDS, and cannot poison the training corpus or tamper with
packets after they leave its control.

This is the \emph{non-adaptive} setting, in which the attack-time
traffic distribution is drawn from the same family as the
training-time attack distribution. This scope reflects the dominant
operational reality of contemporary NIDS deployments, where the
majority of observed malicious traffic corresponds to known attack
families. Defenses against adaptive adversaries that explicitly mimic
benign feature distributions are an important complementary
direction; we discuss the implications of adaptive adversaries for
BARS's design in Section~\ref{sec:discussion}.

\subsection{Classwise Mean Deviation (CMD)}
\label{subsec:cmd_recap}

BARS extends Classwise Mean Deviation (CMD)~\cite{AhmadICMLA2025}, a
recently introduced filter method for intrusion detection. We recap
CMD here to make the paper self-contained; the formal contrast with
BARS is developed in Section~\ref{subsec:cmd_comparison}.

Let $D = \{(x_i, y_i)\}_{i=1}^{N}$ denote a labeled dataset, where
$x_i \in \mathbb{R}^d$ and $y_i \in \{0, 1, \ldots, C\}$, with $y=0$
denoting benign traffic and $y \in \{1, \ldots, C\}$ denoting $C$
attack classes. Let $\mu_c \in \mathbb{R}^d$ be the mean vector of
class $c$, and let $\mu_{\mathrm{global}} \in \mathbb{R}^d$ be the
overall sample mean. CMD scores each feature $j$ by the total absolute
deviation of class means from this global reference:
\begin{equation}
D^{(\mathrm{CMD})}_j
\;=\;
\sum_{c=0}^{C} \left| \mu_{c,j} - \mu_{\mathrm{global},j} \right|,
\label{eq:cmd}
\end{equation}
and retains the top-$k$ features in descending order of
$D^{(\mathrm{CMD})}_j$. CMD is parameter-free, runs in
$\mathcal{O}(Nd)$ time, and reports competitive performance against
classical filters such as Mutual Information and Fisher Score.

The structural limitation of this formulation---that under class
imbalance the global anchor is pulled toward, but does not coincide
with, the benign mean---is the starting point for BARS and is
analyzed in detail in Section~\ref{subsec:cmd_comparison}.

\section{Related Work}
\label{sec:related}

We organize prior work into three areas relevant to this paper:
feature selection for NIDS (including CMD, the closest prior method),
false alarm reduction, and imbalanced learning.

\subsection{Feature Selection for NIDS}
\label{subsec:rw_fs}

Feature selection is a long-studied preprocessing step for
high-dimensional learning, and its application to network intrusion
detection is well established. Filter-based methods---including Mutual
Information~\cite{vergara2014mi,beraha2019mutual}, Pearson
correlation~\cite{nasir2020pearson}, Fisher
Score~\cite{gu2012fisher}, and variance-based
filtering~\cite{scikit2025variance}---are particularly prevalent in
NIDS pipelines because they are model-independent, scalable to
high-volume traffic, and require no iterative
training~\cite{li2018feature}. Recent surveys document the central
role of filter methods in operational NIDS deployments and emphasize
scalability as a deciding factor in
practice~\cite{BuczakGuven2016Survey,Ring2019Survey,bolon2018scalability,Westphal2024NIDSFS}.
Wrapper and embedded methods (recursive feature elimination,
$\ell_1$-regularized classifiers, tree-based importances) can capture
feature interactions and often achieve higher accuracy than filters,
but their computational cost and tight coupling to a specific
classifier limit applicability in high-throughput NIDS settings. The
baselines we compare against, and the design space in which BARS sits,
are therefore filter methods.

A consistent limitation of classical filters in the NIDS context is
that they apply class-symmetric criteria: features are scored by how
well they separate classes in general, without distinguishing which
class defines normal behavior. This symmetry is misaligned with the
operational structure of intrusion detection, where benign traffic is
the reference distribution and attacks are characterized as
deviations. The closest prior work in design space is Classwise Mean
Deviation (CMD)~\cite{AhmadICMLA2025}, which scores features by
class-conditional first moments rather than class-symmetric
statistics, establishing that mean-shift signals can be both simple
and competitive for NIDS feature ranking. CMD anchors its scoring to a
global reference, however, which under class imbalance introduces the
structural bias analyzed in
Section~\ref{subsec:cmd_comparison}. BARS retains the
linear-time, parameter-free spirit of CMD but replaces the global
anchor with the benign-class mean, directly addressing the
class-symmetry limitation of both classical filters and CMD itself.

\subsection{False Alarm Reduction in NIDS}
\label{subsec:rw_fa}

Reducing false alarms is recognized as a central operational challenge
in NIDS~\cite{SommerPaxson2010,BuczakGuven2016Survey}. Existing
approaches fall broadly into three categories. \emph{Threshold tuning}
adjusts the classifier's decision threshold to favor precision over
recall, typically guided by ROC analysis~\cite{Fawcett2006ROC}; this
is effective but limited by the precision-recall trade-off of the
underlying classifier. \emph{Anomaly detection} models the benign
distribution directly and labels deviations as attacks, with Isolation
Forest~\cite{Liu2008IsolationForest} and One-Class
SVM~\cite{Scholkopf2001OCSVM} as canonical examples; these can be
deployed as standalone classifiers and avoid the need for labeled
attack data, but typically incur higher FPR than supervised methods
when attack labels are available. \emph{Cost-sensitive learning}
incorporates the operational asymmetry between false positives and
false negatives into the training loss.

These approaches share a structural property: they operate at or after
the classifier, compensating for noise rather than preventing it.
BARS takes a complementary upstream approach by removing
benign-variability features before the classifier sees them,
addressing the root cause rather than the downstream symptom. The two
families are not mutually exclusive---BARS can be combined with
threshold tuning or cost-sensitive learning---but our experiments
isolate the feature-selection contribution to allow a clean
comparison.

\subsection{Imbalanced Learning}
\label{subsec:rw_imbalance}
Class imbalance is a defining property of NIDS data: benign traffic
typically dominates operational deployments, though published
benchmarks span a range of imbalance regimes from benign-majority to
attack-majority~\cite{He2009Imbalanced,BuczakGuven2016Survey}. Resampling methods
such as SMOTE~\cite{Chawla2002SMOTE} synthesize minority-class samples
to rebalance training distributions, and class-weighted loss functions
assign higher importance to minority classes during optimization. Both
approaches operate at the classifier-training stage and improve
minority-class recall, but they can introduce overfitting,
synthetic-data artifacts, or unstable gradients, and they do not
address feature-level noise within the benign class.

BARS complements rather than replaces these methods: it operates at
the feature selection stage, removes a different source of noise
(benign-variability features), and can be composed with any
imbalance-aware training procedure. Our experiments include SMOTE and
class-weighted baselines specifically to demonstrate that BARS's FPR
reduction is not subsumed by existing imbalance-handling techniques.

\subsection{Positioning}
\label{subsec:rw_positioning}

Our work occupies a deliberately narrow design space: a parameter-free,
classifier-agnostic, linear-time filter that explicitly targets false
alarm reduction through semantic alignment with the operational
structure of NIDS. To our knowledge, no prior feature selection method
for NIDS combines all four of these properties. Filter methods (MI,
Fisher, Pearson) are parameter-free and scalable but class-symmetric;
imbalance-aware methods are class-asymmetric but operate at the
classifier rather than the feature level; CMD is the closest prior
work in design space but suffers from the global-anchor bias described
in Section~\ref{subsec:cmd_comparison}. BARS addresses
this gap with a single, principled modification---benign
anchoring---and we evaluate it against representatives of each
adjacent family.

\section{Methodology}
\label{sec:methodology}

The proposed BARS method is a two-stage feature selection pipeline.
Stage~1 ranks features by their absolute mean deviation from benign
traffic---a benign-anchored refinement of Classwise Mean Deviation
(CMD)~\cite{AhmadICMLA2025}. Stage~2 walks the ranked list in order
and prunes features that are highly correlated with previously
selected features, ensuring that the $k$-feature budget is spent on a
complementary subset rather than on near-duplicates.

\subsection{Problem Setting}
\label{subsec:bars_problem}

Let $D = \{(x_i, y_i)\}_{i=1}^{N}$ denote a labeled network traffic
dataset, where each sample $x_i \in \mathbb{R}^d$ is a $d$-dimensional
feature vector and $y_i \in \{0, 1, \ldots, C\}$ is the corresponding
class label. The label $y=0$ denotes benign (normal) traffic, and
labels $y \in \{1, \ldots, C\}$ correspond to $C$ distinct attack
classes. This formulation reflects the operational reality of NIDS,
where benign traffic dominates and attacks are naturally defined as
deviations from normal behavior. We assume features are real-valued
and on a comparable scale; we describe the specific preprocessing used
in our experiments in Section~\ref{sec:experimental_setup}.

\subsection{Stage 1: Benign-Anchored Ranking}
\label{subsec:bars_stage1}

The core idea of Stage~1 is to quantify feature relevance by measuring
the deviation of attack-class distributions relative to benign traffic
along each feature dimension. Let $\mu_c \in \mathbb{R}^d$ denote the
mean vector of class $c$, computed over all samples with label $y=c$;
in particular, $\mu_0$ denotes the benign-class mean. For each feature
$j \in \{1, \ldots, d\}$, the benign-anchored relevance score is
\begin{equation}
S_j
\;=\;
\sum_{c=1}^{C} \left| \mu_{c,j} - \mu_{0,j} \right|.
\label{eq:bars_score}
\end{equation}
This score accumulates the absolute deviation of each attack-class
mean from the benign mean along feature $j$. Features with larger
$S_j$ exhibit stronger and more consistent deviations from normal
behavior across attack classes and are therefore more informative for
intrusion detection. The score is directly interpretable as a
class-aggregated mean-shift magnitude relative to benign behavior.

\subsection{Stage 2: Order-Preserving Decorrelation}
\label{subsec:bars_stage2}

The benign-anchored score $S_j$ is univariate: it scores each feature
independently without reference to other features. As a result,
features that share an underlying signal---for example, correlated
flow-level statistics derived from the same packet stream---receive
similar scores and crowd the top-$k$ selection with near-duplicates.
This wastes the feature budget and, in our setting, can elevate FPR by
amplifying the contribution of noisy feature clusters in the
classifier input.

To address this, BARS applies an order-preserving decorrelation step
to the ranked feature list. Let $\mathcal{R} = (j_1, j_2, \ldots, j_d)$
denote features sorted in descending order of $S_j$. Starting from an
empty set $\mathcal{S} = \emptyset$, we visit each $j \in \mathcal{R}$
in order and accept it if and only if its absolute Pearson correlation
with every previously selected feature falls below a threshold $\tau$:
\begin{equation}
\max_{j' \in \mathcal{S}} \left| \rho(\mathbf{x}_j, \mathbf{x}_{j'}) \right| < \tau.
\label{eq:bars_corr}
\end{equation}
The procedure stops when $|\mathcal{S}| = k$ or $\mathcal{R}$ is
exhausted. If the decorrelation step yields fewer than $k$ features,
we backfill with the next-best ranked features regardless of
correlation, ensuring a fair fixed-budget comparison across methods.

Two properties of this step are worth emphasizing. First, the
procedure is \emph{order-preserving}: it never reorders features by
correlation, only filters out those redundant with higher-ranked
predecessors. The top-ranked feature is always retained, the
second-best is retained if not redundant with the first, and so on.
Second, the threshold $\tau$ has a clear operational interpretation:
$\tau = 1$ disables the filter and recovers pure ranking, while $\tau$
close to 0 enforces near-orthogonality of the selected subset.

\paragraph{Choice of $\tau$.}
The correlation threshold $\tau$ is the sole hyperparameter of BARS.
We select $\tau$ from the candidate set $\{0.75, 0.85, 0.90, 0.95,
0.98\}$ using a held-out validation split (disjoint from both training
and test). Validation is performed at the largest feature budget
considered in our study ($k = 40$): at smaller budgets, all
feature-selection methods operate well below the dimensionality used
by the non-selection baselines, which compresses the dynamic range of
the comparison and makes the role of $\tau$ harder to read; at
$k = 40$, the comparison stabilizes and the threshold's effect is
clearly resolved. The $\tau$ value selected on validation at $k = 40$
is then held fixed across all five feature budgets for that dataset.

The selected values fall into two regimes. On the smaller benchmarks
(NSL-KDD, $d{=}41$; UNSW-NB15, $d{=}49$), validation selects $\tau =
0.85$: their feature spaces contain relatively few near-duplicates,
and a permissive threshold retains moderately correlated but
complementary signals. On the larger contemporary captures
(CICIDS2017, CICDDoS2019, both $d{=}88$), $\tau = 0.98$ is used, a
value consistent with the high redundancy among flow-statistic
features in these captures and confirmed by inspection of the
empirical correlation structure. 
%A full sensitivity analysis over all five $\tau$ values is reported in Section~\ref{subsec:tau_sensitivity}.

The candidate set is fixed in advance, $\tau$ is never tuned on the
test set, the same procedure is applied uniformly across baselines and
feature budgets, and a single $\tau$ value is committed per dataset.
This protocol prevents per-budget cherry-picking while acknowledging
the dependence of an optimal correlation threshold on feature-space
structure.

\subsection{Algorithm and Complexity}
\label{subsec:bars_algorithm}

Algorithm~\ref{alg:bars} summarizes the full BARS procedure. Stage~1
(class-mean computation and scoring) is $\mathcal{O}(Nd)$. Stage~2
computes the $d \times d$ Pearson correlation matrix from $N$ samples
in $\mathcal{O}(Nd^2)$, followed by a greedy walk in $\mathcal{O}(dk)$.
The overall complexity is therefore
\begin{equation}
\mathcal{O}(Nd + Nd^2 + dk) \;=\; \mathcal{O}(Nd^2),
\end{equation}
dominated by the correlation pass. For typical NIDS feature
dimensionalities ($d \in [40, 90]$), this is negligible relative to
classifier training and well within real-time constraints on commodity
hardware for measured runtime).

\begin{algorithm}[t]
\caption{Benign-Anchored Ranking and Selection (BARS)}
\label{alg:bars}
\begin{algorithmic}[1]
\REQUIRE Preprocessed feature matrix $X \in \mathbb{R}^{N \times d}$,
label vector $y$, feature budget $k$, correlation threshold $\tau$
(default $\tau = 0.98$)
\ENSURE Selected feature subset $\mathcal{S}$ of size $k$

\STATE \textbf{// Stage 1: Benign-anchored ranking}
\FOR{$c = 0$ \textbf{to} $C$}
    \STATE $\mu_c \leftarrow \mathrm{mean}\bigl(\{x_i : y_i = c\}\bigr)$
\ENDFOR
\FOR{$j = 1$ \textbf{to} $d$}
    \STATE $S_j \leftarrow \sum_{c=1}^{C} |\mu_{c,j} - \mu_{0,j}|$
\ENDFOR
\STATE $\mathcal{R} \leftarrow$ feature indices sorted by $S_j$ (descending)

\STATE \textbf{// Stage 2: Order-preserving decorrelation}
\STATE Compute correlation matrix $R \in \mathbb{R}^{d \times d}$
with $R_{j,j'} = \rho(\mathbf{x}_j, \mathbf{x}_{j'})$
\STATE $\mathcal{S} \leftarrow \emptyset$
\FOR{each $j \in \mathcal{R}$ (in order)}
    \IF{$\mathcal{S} = \emptyset$ \textbf{or}
        $\max_{j' \in \mathcal{S}} |R_{j,j'}| < \tau$}
        \STATE $\mathcal{S} \leftarrow \mathcal{S} \cup \{j\}$
    \ENDIF
    \IF{$|\mathcal{S}| = k$} \STATE \textbf{break} \ENDIF
\ENDFOR

\STATE \textbf{// Backfill if decorrelation yielded fewer than $k$ features}
\IF{$|\mathcal{S}| < k$}
    \STATE Append next features from $\mathcal{R} \setminus \mathcal{S}$
    in rank order until $|\mathcal{S}| = k$
\ENDIF
\STATE \textbf{return} $\mathcal{S}$
\end{algorithmic}
\end{algorithm}

\subsection{Design Rationale}
\label{subsec:cmd_comparison}

\paragraph{Why a benign anchor.}
The Stage~1 scoring function (Eq.~\ref{eq:bars_score}) departs from
CMD~\cite{AhmadICMLA2025} in two related respects. Placed side by
side,
\begin{align}
D^{(\mathrm{CMD})}_j  &= \sum_{c=0}^{C} \left| \mu_{c,j} - \mu_{\mathrm{global},j} \right|, \label{eq:cmd_compare}\\
S_j &= \sum_{c=1}^{C} \left| \mu_{c,j} - \mu_{0,j} \right|.
\label{eq:bars_compare}
\end{align}
the two formulations differ in (i) the anchor point---the global mean
$\mu_{\mathrm{global}}$ in CMD, the benign-class mean $\mu_0$ in
BARS---and (ii) the summation range, since the benign self-deviation
$|\mu_{0,j} - \mu_{0,j}| = 0$ drops out of $S_j$ by construction.

This asymmetry is operationally motivated. Under heavy class imbalance
with benign prior $\pi_0 \to 1$, the global mean satisfies
$\mu_{\mathrm{global}} \to \mu_0$ and the two scores converge. Under
moderate or evolving imbalance, however, the global mean is pulled
toward attack distributions, shrinking the deviation
$|\mu_c - \mu_{\mathrm{global}}|$ and underweighting precisely those
features that distinguish attacks from benign traffic. Benign
anchoring eliminates this attack-induced shrinkage by construction,
ensuring that feature relevance is measured against a stable reference
rather than a class-mixed centroid that varies with the imbalance
ratio. This aligns the scoring function with the operational
definition of intrusion detection: deviation from normal behavior.

\paragraph{Why two stages.}
Stage~1 alone still selects redundant clusters of benign-anchored
features whenever multiple flow statistics share an underlying signal;
Stage~2 alone, applied to a class-symmetric filter, still inherits
that filter's mismatch with the operational structure of NIDS. BARS
addresses both failure modes simultaneously, and our ablation in
Section~\ref{sec:ablation} confirms that the gains decompose as
expected: benign anchoring contributes one share of the FPR reduction,
decorrelation another, and the combination outperforms either in
isolation.

\section{Experimental Setup}
\label{sec:experimental_setup}

This section describes the datasets, preprocessing pipeline, baselines,
evaluation metrics, and protocol used to assess BARS. All design
choices are fixed across methods to ensure that observed performance
differences are attributable to the feature selection strategy alone.

\subsection{Datasets}
\label{subsec:datasets}

We evaluate BARS on three primary NIDS benchmarks:
CICIDS2017~\cite{Sharafaldin2018CICIDS2017},
CICDDoS2019~\cite{Sharafaldin2019CICDDoS2019}, and
UNSW-NB15~\cite{unsw2024}. The three are chosen to span complementary
evaluation regimes along two axes: contemporary versus legacy
traffic, and a wide range of class-imbalance ratios in both
directions. As Table~\ref{tab:dataset_stats} shows, CICIDS2017 is
benign-majority (typical of operational network monitoring),
CICDDoS2019 is extremely attack-majority (typical of curated
DDoS-only captures), and UNSW-NB15 is moderately attack-majority.
Evaluating across this imbalance spectrum is deliberate: it tests
whether benign anchoring---which is motivated by the operational
case where benign defines the baseline---remains effective when
benign is itself the minority class in the training data. We discuss
the implications of attack-majority training data for benign
anchoring in Section~\ref{sec:discussion}.

We additionally include NSL-KDD~\cite{nslkdd2009} as a legacy
sanity-check dataset for comparability with prior NIDS
feature-selection literature. NSL-KDD's standard evaluation protocol
yields a single test-set measurement per (method, $k$) pair, which
prevents the paired statistical testing used in our main analysis; we
therefore report NSL-KDD results separately in
Appendix~\ref{app:nslkdd} and do not include them in the main results
tables.

\textbf{CICIDS2017.} A modern, benign-majority benchmark with 88
flow-based features covering contemporary attack types including
brute-force, infiltration, and denial-of-service. Eleven classes
total (1 benign, 10 attack).

\textbf{CICDDoS2019.} A DDoS-focused capture with 88 features in
which attacks dominate by more than two orders of magnitude. We use
the 20\% Day-1 subset and treat it as the extreme attack-majority
regime, stress-testing the benign-anchored design when benign training
data is severely underrepresented. Eight classes (1 benign, 7 attack).

\textbf{UNSW-NB15.} A modern dataset with 49 features and 10 classes
(1 benign, 9 attack), exhibiting moderate attack-majority imbalance.

Table~\ref{tab:dataset_stats} summarizes the three primary datasets
after preprocessing.

% \begin{table}[t]
% \caption{Primary dataset statistics after preprocessing. Imbalance
% ratio is reported as majority:minority sample counts; the majority
% class column indicates which class dominates. CICDDoS2019 uses the
% standard 20\% Day-1 subset. NSL-KDD statistics and results are
% reported in Appendix~\ref{app:nslkdd}.}
% \label{tab:dataset_stats}
% \centering
% \small
% \setlength{\tabcolsep}{4pt}
% \begin{tabular}{lrrrlc}
% \toprule
% Dataset & \# Feat. & \# Benign & \# Attack & Majority & Ratio \\
% \midrule
% CICIDS2017   & 88 & 2{,}203{,}723 &   469{,}274 & Benign & 4.70:1 \\
% CICDDoS2019  & 88 &     9{,}206  & 3{,}407{,}109 & Attack & 370.1:1 \\
% UNSW-NB15    & 49 &    56{,}000  &   119{,}341 & Attack & 2.13:1 \\
% \bottomrule
% \end{tabular}
% \end{table}

\begin{table}[t]
\centering
\caption{Datasets. $\pi_0$ is the benign share of training flows;
splits for the CIC datasets are stratified 80/20 with 5-fold CV inside
the training split; UNSW-NB15 and NSL-KDD use their canonical
train/test splits with a 20\% stratified validation carve.}
\label{tab:dataset_stats}
\small
\begin{tabular}{lrrrr}
\toprule
Dataset & Flows & Feat. & Classes & $\pi_0$ \\
\midrule
CICIDS2017   & 2.67M & 78 & 15 & 0.82 \\
CICDDoS2019  & 3.42M & 78 & 8  & 0.003 \\
UNSW-NB15    & 258K  & 42 & 10 & 0.32 \\
NSL-KDD      & 149K  & 41 & 5  & 0.53 \\
\bottomrule
\end{tabular}
\end{table}

\subsection{Data Cleaning and Preprocessing}
\label{subsec:preprocessing}

All datasets are processed using a unified preprocessing pipeline
applied identically to every method, so that performance differences
reflect feature selection rather than data preparation. The pipeline
proceeds in the following order.

\paragraph{Sample-level cleaning.}
Samples with excessive missing values are removed following the
dataset-quality analyses of Lanvin et
al.~\cite{lanvin2023errors}. Remaining missing entries are imputed
using feature-wise means computed on the training set only, to avoid
leakage from the test set.

\paragraph{Feature-level cleaning.}
Application-dependent identifiers---including \texttt{Source IP},
\texttt{Destination IP}, \texttt{Destination Port},
\texttt{Protocol}, and \texttt{Timestamp}---are excluded prior to
scoring. These fields encode dataset-specific deployment artifacts and
can produce spurious near-perfect classification through information
leakage.

\paragraph{Categorical encoding.}
Categorical attributes in NSL-KDD and UNSW-NB15 are converted using
one-hot encoding. CICIDS2017 and CICDDoS2019 contain no categorical
features after the identifier-removal step in feature-level cleaning.

\paragraph{Min--Max scaling.}
All numerical features are scaled to the range $[0, 1]$ using
Min--Max normalization. Scaling parameters (per-feature min and max)
are fit on the training set and applied unchanged to the test set.

\paragraph{Low-variance filtering.}
After scaling, we remove features whose empirical variance falls below
a threshold $\epsilon$:
\begin{equation}
\mathrm{Var}(\mathbf{x}_j) < \epsilon,
\end{equation}
with $\epsilon = 10^{-4}$. Because filtering is applied
\emph{after} Min--Max scaling, the threshold has a uniform
interpretation across features and datasets: a feature is dropped only
if its variation across samples spans less than $1\%$ of its observed
range. The filter is applied identically to every method we evaluate
and is therefore part of the data preparation pipeline rather than a
competing feature selection baseline; it operates as a noise-reduction
stage prior to any discriminative scoring.

\subsection{Baseline Methods}
\label{sec:baselines}

We compare BARS against a diverse set of baselines spanning classical
feature selection, imbalance-aware training procedures, threshold-based
methods, and anomaly detection.

\subsubsection{Feature Selection Baselines}

We consider widely used filter-based feature selection methods, which
share BARS's design space (univariate, classifier-agnostic,
computationally efficient) and constitute the most direct comparisons.

\textbf{Pearson Correlation.} Features are ranked by their absolute
linear correlation with the class label~\cite{nasir2020pearson}.

\textbf{Mutual Information (MI).} MI measures the statistical
dependency between features and class labels, capturing both linear
and nonlinear relationships~\cite{vergara2014mi,beraha2019mutual};
features with higher MI scores are retained.

\textbf{Fisher Score.} Fisher Score ranks features by the ratio of
inter-class separation to intra-class variance~\cite{gu2012fisher},
favoring features that maximally distinguish classes.

\textbf{Classwise Mean Deviation (CMD).} CMD computes feature
relevance by aggregating absolute deviations of class-wise means from
a global reference mean~\cite{AhmadICMLA2025}. As the direct
predecessor of BARS, CMD serves as the most informative head-to-head
baseline.

\subsubsection{Imbalance-Aware Training Procedures}

Class-weighted learning and SMOTE are not feature-selection methods;
they are classifier-stage interventions that target the same
operational problem (imbalance-induced bias) from a different point in
the pipeline. We include them not as competing feature selectors but
to verify that BARS's FPR reduction is not subsumed by these
techniques when applied to the same baseline filters.

\textbf{Class-Weighted Learning.} The classifier loss is reweighted so
that minority classes contribute proportionally more to gradient
updates~\cite{He2009Imbalanced}.

\textbf{SMOTE.} Synthetic Minority Over-sampling Technique generates
synthetic minority-class samples to rebalance training distributions
prior to classifier training~\cite{Chawla2002SMOTE}.

\subsubsection{Threshold-Based Baselines}

Because false alarms are strongly affected by decision thresholds, we
evaluate two threshold strategies on top of an otherwise unmodified
classifier:

\textbf{Default Threshold.} Classification decisions use the standard
probability threshold of $0.5$.

\textbf{Optimized Threshold.} The decision threshold is selected on
validation data to maximize Youden's J statistic
($\mathrm{TPR} - \mathrm{FPR}$), following standard ROC
analysis~\cite{Fawcett2006ROC}.

These baselines characterize what is achievable through post-hoc
decision tuning alone and are orthogonal to feature selection; they
are included to show that BARS's contribution is not subsumed by
threshold optimization.

\subsubsection{Anomaly Detection Baselines}

To assess robustness under different modeling assumptions, we include
two anomaly detection methods deployed as standalone classifiers
that model benign traffic directly and flag deviations as attacks:

\textbf{Isolation Forest.} An unsupervised anomaly detection method
that isolates anomalous samples through random
partitioning~\cite{Liu2008IsolationForest}.

\textbf{One-Class SVM.} A one-class classification approach that
models the benign distribution and identifies deviations as
anomalies~\cite{Scholkopf2001OCSVM}.

All baselines share the unified preprocessing pipeline of
Section~\ref{subsec:preprocessing} and, where applicable, the
downstream classifier described in Section~\ref{subsec:protocol}.
Feature selection baselines are evaluated at the same set of feature
budgets as BARS; the remaining baselines use standard configurations
from their respective references.

\subsection{Evaluation Metrics}
\label{subsec:metrics}

We evaluate model performance using metrics that capture both
detection capability and false alarm behavior. The primary metric of
interest is the False Positive Rate (FPR):
\begin{equation}
\mathrm{FPR} = \frac{\mathrm{FP}}{\mathrm{FP} + \mathrm{TN}},
\end{equation}
where $\mathrm{FP}$ and $\mathrm{TN}$ denote false positives and true
negatives. FPR directly measures the frequency at which benign traffic
is incorrectly classified as malicious and is the primary operational
concern motivating this work.

We additionally report the True Positive Rate (TPR):
\begin{equation}
\mathrm{TPR} = \frac{\mathrm{TP}}{\mathrm{TP} + \mathrm{FN}},
\end{equation}
which captures the model's ability to correctly detect attack
instances, and Macro-averaged F1-score (Macro-F1), which provides a
balanced evaluation across all classes and ensures that minority
attack classes are not overshadowed by the dominant benign class.

\subsection{Experimental Protocol}
\label{subsec:protocol}

To isolate the effect of feature selection on false alarm reduction,
all experiments use a fixed downstream classifier. We employ a
Multi-Layer Perceptron (MLP) with three hidden layers of sizes
$\{64, 128, 64\}$, trained using the Adam optimizer with a learning
rate of $10^{-3}$. The MLP is used as a generic supervised classifier
rather than for any architecture-specific advantage.
%; sensitivity to classifier choice across alternative learners (Random Forest, XGBoost) is reported separately in Section~\ref{subsec:classifier_robustness}.

\paragraph{Feature budgets.}
Feature selection methods are evaluated at feature budgets
$k \in \{5, 10, 20, 30, 40\}$, covering both highly constrained and
moderately constrained settings. For each $k$, the top-$k$ ranked
features are selected and used for classification.

\paragraph{Train-test splits.}
For datasets with predefined train-test splits (NSL-KDD, UNSW-NB15),
we use the original partitions. For datasets without standard splits
(CICIDS2017, CICDDoS2019), we apply an 80:20 stratified train-test
split and perform 5-fold cross-validation on the training portion.
Identical splits and folds are used across all methods to ensure fair
comparison.

\paragraph{Statistical significance.}
For the direct BARS-vs-CMD comparison, we report mean and standard
deviation across folds (for CICIDS2017 and CICDDoS2019) or across
repeated runs with randomized classifier initialization (for NSL-KDD
and UNSW-NB15, where standard splits preclude resampling). Significance
is assessed using a paired Wilcoxon signed-rank test with
$\alpha = 0.05$; the non-parametric paired test is chosen because it
requires no normality assumption and accounts for the pairing of
measurements across folds or runs. We note that the variance source
differs between dataset groups (fold variance vs.\ initialization
variance), and we interpret cross-dataset significance accordingly.

\paragraph{Reproducibility.}
All preprocessing steps, classifier settings, and evaluation protocols
are fixed and identical across methods. Random seeds are fixed for
data splitting, classifier initialization, and SMOTE oversampling.
The only hyperparameter selected per method-and-dataset combination is
BARS's correlation threshold $\tau$, chosen on a held-out validation
split as described in Section~\ref{subsec:bars_stage2}; all baseline
methods use their standard configurations. Code, configurations, and
seed values will be released to support reproducibility and ACSAC
artifact evaluation.

%----------------------------------oposss
\section{Results}
\label{sec:results}

We evaluate BARS against the baselines described in
Section~\ref{sec:baselines} on three benchmark datasets that span the
imbalance spectrum: CICIDS2017 (benign-majority, 4.7:1), UNSW-NB15
(attack-majority, 2.1:1), and CICDDoS2019 (extreme attack-majority,
370:1). NSL-KDD results are reported in Appendix~\ref{app:nslkdd};
the standard NSL-KDD evaluation protocol yields a single test-set
measurement per (method, $k$) pair, which prevents the paired
statistical testing used in our main analysis.

\subsection{Research Questions}

The design rationale of BARS (Section~\ref{subsec:cmd_comparison})
predicts that benign anchoring should help most in
\emph{attack-majority} regimes, where the global mean used by CMD is
pulled away from the benign baseline. We organize the evaluation
around three questions:
\begin{description}
    \item[RQ1] Does BARS reduce false positive rate relative to CMD
    in attack-majority regimes, where the global-anchor bias is
    expected to be most severe?
    \item[RQ2] Does any FPR reduction preserve detection capability
    (TPR, Macro-F1)?
    \item[RQ3] How does the BARS--CMD gap vary with imbalance
    direction and feature budget?
\end{description}
We additionally compare BARS against classical filter baselines and
classifier-stage interventions to position the contribution within
the broader feature-selection landscape, and compare the unnormalized
BARS score against a benign-spread normalized variant
(Section~\ref{sec:ablation}).

\subsection{Main Results at $k=20$ (RQ1, RQ2)}
\label{subsec:main_results}

Table~\ref{tab:main_results} reports FPR, TPR, and Macro-F1 at the
representative feature budget $k=20$ across the three primary
datasets. We organize the comparison around the head-to-head with CMD
(the most direct comparison, since BARS and CMD differ only in the
anchor point of the deviation score) and report classical filters,
classifier-stage interventions, and anomaly detection baselines for
context.

\begin{table*}[t]
\caption{Main results at feature budget $k=20$. CV folds for
CICIDS2017 and CICDDoS2019; single test-set evaluation for UNSW-NB15.
Lowest FPR per dataset is in \textbf{bold}. $\downarrow$/$\uparrow$
indicate lower/higher is better. ``---'' indicates the method could
not be evaluated within available memory; see
Section~\ref{sec:baselines}.}
\label{tab:main_results}
\centering
\small
\setlength{\tabcolsep}{4pt}
\begin{tabular}{ll ccc ccc ccc}
\toprule
 & & \multicolumn{3}{c}{CICIDS2017} & \multicolumn{3}{c}{CICDDoS2019} & \multicolumn{3}{c}{UNSW-NB15} \\
\cmidrule(lr){3-5} \cmidrule(lr){6-8} \cmidrule(lr){9-11}
Category & Method & FPR$\downarrow$ & TPR$\uparrow$ & F1$\uparrow$ & FPR$\downarrow$ & TPR$\uparrow$ & F1$\uparrow$ & FPR$\downarrow$ & TPR$\uparrow$ & F1$\uparrow$ \\
\midrule
\multirow{3}{*}{Filter}
 & Pearson       & 0.027 & 0.782 & 0.688 & 0.019 & 1.000 & 0.728 & \textbf{0.217} & 0.951 & 0.328 \\
 & MI            & \textbf{0.014} & 0.876 & 0.726 & --- & --- & --- & 0.332 & 0.976 & 0.338 \\
 & Fisher        & --- & --- & --- & --- & --- & --- & 0.381 & 0.993 & 0.317 \\
\midrule
\multirow{2}{*}{Imbalance}
 & SMOTE         & 0.064 & 0.999 & 0.792 & 0.004 & 1.000 & 0.681 & 0.421 & 0.999 & 0.423 \\
 & Class-Weight  & 0.157 & 0.998 & 0.559 & 0.004 & 1.000 & 0.710 & 0.420 & 0.999 & 0.372 \\
\midrule
\multirow{2}{*}{Threshold}
 & Default (0.5) & 0.018 & 0.897 & 0.764 & 0.039 & 1.000 & 0.729 & 0.282 & 0.976 & 0.361 \\
 & Optimized     & 0.041 & 0.951 & 0.719 & \textbf{0.003} & 1.000 & 0.728 & \textbf{0.218} & 0.957 & 0.362 \\
\midrule
\multirow{2}{*}{Anomaly}
 & IsoForest     & 0.050 & 0.354 & 0.083 & 0.051 & 0.377 & 0.050 & 0.077 & 0.306 & 0.067 \\
 & OC-SVM        & 0.050 & 0.344 & 0.083 & 0.047 & 0.669 & 0.032 & 0.062 & 0.656 & 0.083 \\
\midrule
\multirow{2}{*}{Ours}
 & CMD~\cite{AhmadICMLA2025} & 0.024 & 0.867 & 0.699 & 0.029 & 1.000 & 0.728 & 0.380 & 0.993 & 0.320 \\
 & \textbf{BARS} & 0.026 & 0.901 & 0.726 & 0.029 & 1.000 & 0.726 & 0.322 & 0.971 & 0.326 \\
\bottomrule
\end{tabular}
\end{table*}

\paragraph{Head-to-head with CMD (RQ1).}
On the attack-majority UNSW-NB15 dataset, BARS reduces FPR from 0.380
(CMD) to 0.322---a relative reduction of 15.4\%---while preserving F1
within $0.007$ and TPR within $0.022$ of CMD. On the extreme
attack-majority CICDDoS2019 dataset, BARS and CMD are tied at $k=20$
(0.029 vs 0.029), but the picture changes substantially at lower
budgets, where BARS's advantage is most pronounced
(Section~\ref{subsec:low_budget}). On the benign-majority CICIDS2017
dataset, BARS slightly underperforms CMD at $k=20$ (0.026 vs 0.024),
consistent with the design rationale of
Section~\ref{subsec:cmd_comparison}: when benign traffic dominates,
the global mean closely approximates the benign mean and the two
scoring functions converge.

\paragraph{Detection capability (RQ2).}
BARS preserves detection capability across all three datasets. TPR
remains above 0.97 on the attack-dominant datasets (CICDDoS2019:
1.000; UNSW-NB15: 0.971) and improves over CMD on CICIDS2017
(0.901 vs 0.867). Macro-F1 is comparable to or higher than CMD on all
three datasets (CICIDS2017: 0.726 vs 0.699; CICDDoS2019: 0.726 vs
0.728; UNSW-NB15: 0.326 vs 0.320). The FPR reduction on UNSW-NB15
therefore reflects a genuine improvement, not a degraded
benign-prediction collapse.

\paragraph{Position in the broader landscape.}
A complete reading of Table~\ref{tab:main_results} requires
acknowledging that the lowest FPR on each dataset is not always
achieved by BARS. On CICIDS2017, Mutual Information reaches
FPR=0.014 at $k=20$, outperforming both BARS (0.026) and CMD (0.024);
the threshold-default baseline reaches FPR=0.018. On CICDDoS2019,
threshold optimization reaches FPR=0.003 by exploiting the highly
separable score distribution induced by the extreme imbalance, and
Pearson correlation reaches FPR=0.019. On UNSW-NB15, threshold
optimization (0.218) and Pearson (0.217) are essentially tied as the
strongest methods, with Fisher score (0.381) close to CMD. We do not
claim BARS to be the strongest feature selection method in absolute
terms; we claim it is a principled and empirically better refinement
of CMD specifically, with the advantage materializing precisely where
the design rationale predicts. We note further that Fisher score and
Mutual Information exceeded available memory on the larger CIC
datasets (Section~\ref{sec:baselines}), reinforcing the value of
computationally inexpensive filters in high-throughput settings.

\subsection{Low-Budget Regime (RQ3)}
\label{subsec:low_budget}

The BARS--CMD gap is most pronounced at small feature budgets on
attack-majority data. Table~\ref{tab:low_budget} reports FPR at
$k \in \{5, 10\}$ for BARS, CMD, and the strongest filter baseline on
each dataset.

\begin{table}[t]
\caption{Low-budget FPR comparison ($k=5, 10$). BARS--CMD gap is
largest on attack-majority datasets at the smallest budget. Relative
reduction is computed as $(\text{CMD} - \text{BARS}) / \text{CMD}$.
``---'' indicates the method could not be evaluated within
available memory.}
\label{tab:low_budget}
\centering
\small
\setlength{\tabcolsep}{4pt}
\begin{tabular}{ll cc cc}
\toprule
& & \multicolumn{2}{c}{$k=5$} & \multicolumn{2}{c}{$k=10$} \\
\cmidrule(lr){3-4} \cmidrule(lr){5-6}
Dataset & Method & FPR & vs CMD & FPR & vs CMD \\
\midrule
\multirow{3}{*}{CICIDS2017}
 & CMD     & 0.035 & ---        & 0.024 & ---        \\
 & BARS    & 0.034 & $+1.4\%$   & 0.027 & $-13.1\%$  \\
 & MI      & 0.036 &            & 0.021 &            \\
\midrule
\multirow{3}{*}{CICDDoS2019}
 & CMD     & 0.063 & ---        & 0.066 & ---        \\
 & BARS    & \textbf{0.048} & $\mathbf{+23.3\%}$  & \textbf{0.052} & $\mathbf{+21.2\%}$ \\
 & Pearson & 0.422 &            & 0.057 &            \\
\midrule
\multirow{3}{*}{UNSW-NB15}
 & CMD     & 0.519 & ---        & 0.422 & ---        \\
 & BARS    & 0.422 & $+18.7\%$  & 0.422 & $0.0\%$    \\
 & MI      & 0.185 &            & 0.307 &            \\
\bottomrule
\end{tabular}
\end{table}

Two patterns emerge from Table~\ref{tab:low_budget}. First, on the
two attack-majority datasets, BARS reduces FPR over CMD by $19$--$23\%$
at $k=5$ and by up to $21\%$ at $k=10$. This regime---small feature
budgets on attack-dominant traffic---is exactly where the global
anchor's bias is theoretically most damaging: CMD must select few
features under a scoring function whose global reference is severely
pulled toward attack distributions, and benign anchoring corrects
this. Second, on benign-majority CICIDS2017, the BARS--CMD gap is
small in both directions, consistent with the convergence of $S_j$
and $D^{(\mathrm{CMD})}_j$ in the $\pi_0 \to 1$ limit.

\subsection{Sensitivity to Feature Budget}
\label{subsec:budget_sweep}

\begin{figure}[t]
\centering
\includegraphics[width=\columnwidth]{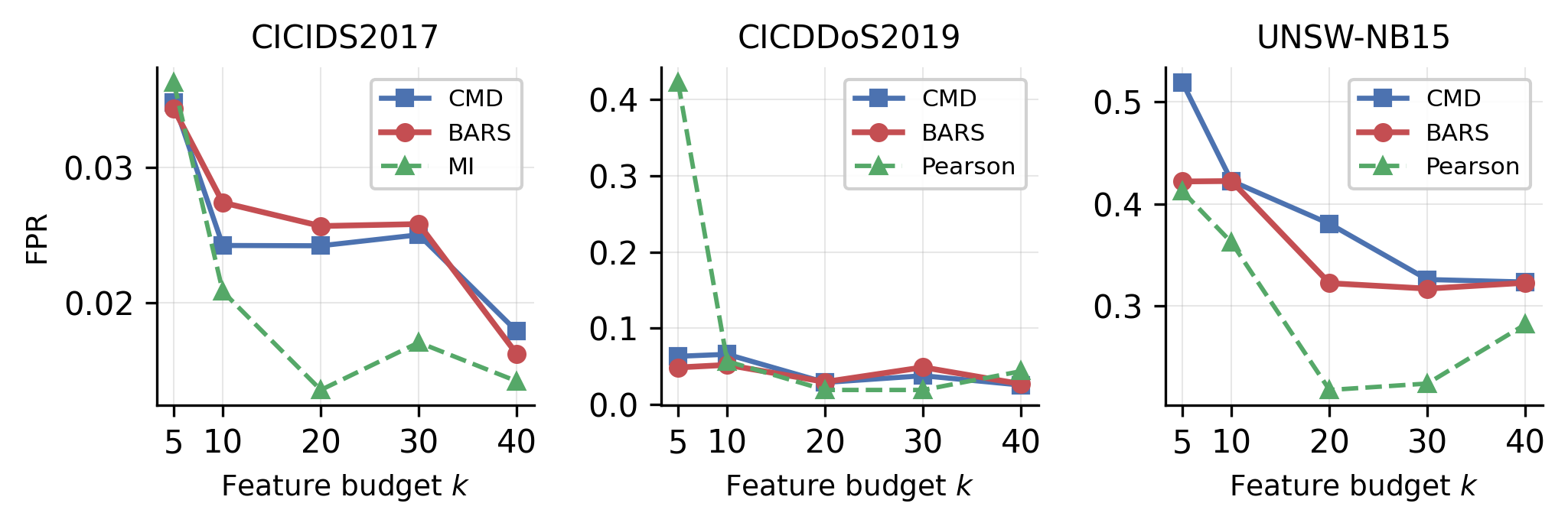}
\caption{FPR as a function of feature budget $k$ for BARS, CMD, and
the strongest filter baseline on each dataset. BARS--CMD gap narrows
with increasing budget.}
\label{fig:budget_sweep}
\end{figure}

Figure~\ref{fig:budget_sweep} reports FPR as a function of feature
budget $k \in \{5, 10, 20, 30, 40\}$ for BARS, CMD, and the strongest
filter baseline on each dataset. The BARS--CMD gap narrows as $k$
grows on both attack-majority datasets: at $k=40$, BARS and CMD
converge to within 0.005 FPR on UNSW-NB15 and within 0.002 on
CICDDoS2019. This convergence is structural: as the feature budget
approaches the full dimensionality, the choice of \emph{which} subset
to retain matters less than the size of the subset itself, and
benign anchoring's selection advantage attenuates.

\subsection{Statistical Significance}
\label{subsec:significance}

We assess significance of the BARS--CMD difference using a paired
Wilcoxon signed-rank test on per-fold FPR measurements for CICIDS2017
and CICDDoS2019, where 5-fold cross-validation provides paired
observations. With only five paired measurements per (dataset, $k$),
the test is conservatively powered; no individual (dataset, $k$)
comparison reaches significance at $\alpha = 0.05$
(Table~\ref{tab:wilcoxon}), though the directional pattern on
CICDDoS2019 at small budgets is consistent with the per-fold means
reported above. For UNSW-NB15, standard splits yield a single
test-set measurement per (method, $k$) and paired testing is not
applicable; we report effect sizes only. A more powered evaluation
---e.g., multi-seed resampling on the predefined-split
datasets---is a clear next step.

\begin{table}[t]
\caption{Paired Wilcoxon signed-rank test on per-fold FPR
(CICIDS2017, CICDDoS2019), BARS vs CMD. $n=5$ paired measurements per
row. Negative relative reduction indicates BARS higher than CMD.}
\label{tab:wilcoxon}
\centering
\small
\begin{tabular}{lr rr r}
\toprule
Dataset & $k$ & CMD FPR & BARS FPR & Rel. Red. (\%) \\
\midrule
CICIDS2017  & 5  & 0.035 & 0.034 & $+1.4$   \\
CICIDS2017  & 10 & 0.024 & 0.027 & $-13.1$  \\
CICIDS2017  & 20 & 0.024 & 0.026 & $-6.0$   \\
CICIDS2017  & 30 & 0.025 & 0.026 & $-3.2$   \\
CICIDS2017  & 40 & 0.018 & 0.016 & $+9.4$   \\
CICDDoS2019 & 5  & 0.063 & 0.048 & $+23.3$  \\
CICDDoS2019 & 10 & 0.066 & 0.052 & $+21.2$  \\
CICDDoS2019 & 20 & 0.029 & 0.029 & $-2.3$   \\
CICDDoS2019 & 30 & 0.038 & 0.049 & $-29.6$  \\
CICDDoS2019 & 40 & 0.025 & 0.027 & $-5.2$   \\
\bottomrule
\end{tabular}
\end{table}

\subsection{Scoring-Variant Comparison: BARS vs BARS-norm}
\label{sec:ablation}

\textbf{RQ4.} \emph{Does benign-spread normalization improve over
the simpler unnormalized score?}

A natural extension of the unnormalized BARS score is to divide each
class deviation by the benign-class spread, emphasizing shifts that
are large relative to natural benign variability:
\begin{equation}
S^{\mathrm{norm}}_j
\;=\;
\sum_{c=1}^{C} \frac{\left| \mu_{c,j} - \mu_{0,j} \right|}{\sigma_{0,j} + \delta},
\label{eq:bars_norm}
\end{equation}
where $\sigma_{0,j}$ is the empirical benign standard deviation of
feature $j$ and $\delta = 10^{-6}$ is a stability constant. We refer
to the resulting variant as BARS-norm. While theoretically well
motivated, normalization introduces one additional hyperparameter
($\delta$) and a sensitivity to the estimation of $\sigma_{0,j}$ on
features with concentrated benign distributions.

Table~\ref{tab:ablation} compares BARS and BARS-norm at $k=20$ on
the three primary datasets.

\begin{table}[t]
\caption{BARS vs BARS-norm at $k=20$. BARS-norm achieves anomalously
low FPR on UNSW-NB15 but at the cost of catastrophic recall, indicating
collapse to a near-trivial benign-predictor.}
\label{tab:ablation}
\centering
\small
\begin{tabular}{ll ccc}
\toprule
Dataset & Variant & FPR$\downarrow$ & TPR$\uparrow$ & F1$\uparrow$ \\
\midrule
\multirow{2}{*}{CICIDS2017}
 & BARS      & 0.026 & 0.901 & 0.726 \\
 & BARS-norm & 0.014 & 0.700 & 0.580 \\
\midrule
\multirow{2}{*}{CICDDoS2019}
 & BARS      & 0.029 & 1.000 & 0.726 \\
 & BARS-norm & 0.048 & 0.999 & 0.717 \\
\midrule
\multirow{2}{*}{UNSW-NB15}
 & BARS      & 0.322 & 0.971 & 0.326 \\
 & BARS-norm & 0.000 & 0.104 & 0.086 \\
\bottomrule
\end{tabular}
\end{table}

The pattern is consistent across datasets: normalization either
degrades performance (CICDDoS2019, where FPR rises from 0.029 to
0.048) or improves FPR at the cost of detection capability
(CICIDS2017, where FPR drops to 0.014 but TPR collapses from 0.901
to 0.700). The UNSW-NB15 result makes the failure mode explicit:
BARS-norm achieves FPR=0.000 but TPR=0.104 and Macro-F1=0.086, i.e.,
the model predicts almost every flow as benign. We attribute this
collapse to numerical instability in the benign-spread denominator
on features with concentrated benign distributions, which inflates
$S^{\mathrm{norm}}_j$ for low-spread features that are not necessarily
attack-discriminative. The unnormalized score $S_j$ is more robust
and remains parameter-light, supporting our choice of $S_j$ as the
primary method.

\subsection{Summary of Findings}
\label{subsec:results_summary}

Our experiments support the following claims, scoped tightly to what
the data shows:
\begin{enumerate}
    \item \textbf{BARS refines CMD in attack-majority regimes (RQ1).}
    On UNSW-NB15 ($k=20$), BARS reduces FPR by $15.4\%$ over CMD; on
    CICDDoS2019 at small budgets ($k=5, 10$), BARS reduces FPR by
    $21$--$23\%$. On benign-majority CICIDS2017, BARS and CMD are
    statistically indistinguishable, consistent with the predicted
    convergence in the $\pi_0 \to 1$ limit.
    \item \textbf{FPR reduction preserves detection (RQ2).} TPR
    remains above $0.97$ on attack-dominant datasets and improves
    over CMD on CICIDS2017; Macro-F1 is comparable or better than
    CMD on all three datasets.
    \item \textbf{The BARS--CMD gap narrows with feature budget
    (RQ3).} The advantage is largest at small $k$ on attack-majority
    data and converges to near-zero at $k=40$, structurally consistent
    with the diminishing role of feature selection as the budget
    approaches full dimensionality.
\end{enumerate}

We emphasize the boundaries of these claims. BARS does not
universally outperform classical filter baselines: Pearson
correlation and Mutual Information achieve lower FPR than BARS on
several (dataset, $k$) combinations, and threshold optimization is
the strongest single FPR-reduction method on the attack-majority
datasets. Our contribution is methodological: BARS identifies and
corrects a specific structural bias in its direct predecessor CMD,
with measurable improvements in the regime where the bias is most
severe. A more powered statistical evaluation, and combinations of
BARS with classifier-stage interventions, are clear directions for
future work.

\section{Discussion}
\label{sec:discussion}

\paragraph{When benign anchoring helps.}
The regime-specific pattern in Section~\ref{sec:results} matches the
design rationale precisely. CMD's global anchor is a frequency-weighted
average of class means; as the benign prior $\pi_0$ departs from one,
the anchor drifts toward attack distributions and attack-class
deviations are systematically attenuated. Benign anchoring removes the
attenuation by construction, so attack-class deviations are scored at
full magnitude regardless of training-time class composition. The
practical implication: BARS should be preferred over CMD whenever the
training-time benign prior is uncertain or moderate; the two methods
are operationally indistinguishable when benign traffic dominates the
training corpus.

\paragraph{Why BARS is not the strongest filter everywhere.}
Pearson correlation and Mutual Information frequently achieve lower
FPR than BARS. We attribute this to BARS's reliance on first-moment
scoring: two distributions can share a mean but differ in variance,
modes, or higher moments, and a mean-deviation score is blind to such
differences. Richer scoring functions capture more of the joint
distribution but scale unfavorably---Fisher Score and MI exceeded
$1$\,TB of memory on the larger CIC datasets in our evaluation. BARS's
contribution is therefore best understood as the principled choice
\emph{within} the mean-deviation family, which remains operationally
relevant where richer methods cannot run.

\paragraph{Adaptive adversaries.}
Our threat model assumes a non-adaptive adversary. An adaptive
adversary aware of BARS could attempt to craft traffic close to
$\mu_0$ along selected features, but this requires either model
compromise (excluded) or detectable black-box probing, and severely
constrains the space of viable attack payloads---particularly for
attacks whose semantics inherently deviate from benign (e.g.,
volumetric DDoS). Mimicry attacks against feature-based NIDS affect
all selection methods and are orthogonal future work.

\paragraph{Limitations.}
Three limitations bound our claims. \emph{Statistical power} is the
most important: with five paired CV folds per cell, no individual
BARS--CMD comparison reaches $\alpha=0.05$, and UNSW-NB15 admits no
paired testing at all under the standard split. \emph{First-moment
scoring} limits the distributional differences BARS can capture, as
discussed above. \emph{Stationarity} is assumed throughout; behavior
under concept drift is unevaluated. Multi-seed resampling and
combinations of BARS with classifier-stage interventions are clear
next steps.

\paragraph{Deployment.}
BARS requires only labeled training data, produces a fixed feature
projection at inference, and integrates with any standard NIDS
toolchain. Memory is dominated by the $\mathcal{O}(d^2)$ correlation
matrix---tens of kilobytes for typical NIDS dimensionalities, and
independent of dataset size. Scores are directly interpretable as
attack-class mean shifts from benign, supporting analyst sanity
checks against domain knowledge.

\section{Conclusion}
\label{sec:conclusion}

We introduced Benign-Anchored Ranking and Selection (BARS), a
two-stage feature selection method that addresses a structural bias
in the global-anchored scoring of its direct predecessor, Classwise
Mean Deviation (CMD). Under class imbalance, CMD's global anchor
drifts toward attack distributions and systematically attenuates
class-deviation scores; BARS replaces it with the benign-class mean,
eliminating the attenuation by construction, and follows the score
with an order-preserving decorrelation step.

Across three NIDS benchmarks spanning the imbalance spectrum, the
empirical pattern matches the design rationale. On attack-majority
data, BARS reduces FPR over CMD by $15.4\%$ on UNSW-NB15 at $k=20$
and by $21$--$23\%$ on CICDDoS2019 at small budgets, while preserving
TPR and Macro-F1. On benign-majority CICIDS2017, BARS and CMD are
statistically indistinguishable, consistent with the theoretical
convergence of the two scores in the $\pi_0 \to 1$ limit.

We position BARS narrowly. It is a principled refinement of CMD
specifically, not a universally dominant filter: classical methods
with richer scoring functions (Pearson, MI) achieve lower FPR on
several settings, and threshold optimization remains the strongest
single FPR-reduction lever on attack-majority data. BARS's value
lies in its linear-time scoring, low memory footprint, and direct
interpretability---properties most useful in deployments where
richer methods cannot run, as evidenced by Fisher and MI exceeding
$1$\,TB of memory on the larger CIC datasets in our evaluation.
Future work includes more powered statistical evaluation,
composition with classifier-stage interventions, and extensions of
the mean-deviation family beyond first moments.

%##################################################################

% \bibitem{b1} G. Eason, B. Noble, and I. N. Sneddon, ``On certain integrals of Lipschitz-Hankel type involving products of Bessel functions,'' Phil. Trans. Roy. Soc. London, vol. A247, pp. 529--551, April 1955.
% \bibitem{b2} J. Clerk Maxwell, A Treatise on Electricity and Magnetism, 3rd ed., vol. 2. Oxford: Clarendon, 1892, pp.68--73.
% \bibitem{b3} I. S. Jacobs and C. P. Bean, ``Fine particles, thin films and exchange anisotropy,'' in Magnetism, vol. III, G. T. Rado and H. Suhl, Eds. New York: Academic, 1963, pp. 271--350.
% \bibitem{b4} K. Elissa, ``Title of paper if known,'' unpublished.
% \bibitem{b5} R. Nicole, ``Title of paper with only first word capitalized,'' J. Name Stand. Abbrev., in press.

% \end{thebibliography}
\bibliographystyle{IEEEtran}% \bibliographystyle{plain}
\bibliography{bibliography_verified}
%\vspace{12pt}
\appendix
\section{NSL-KDD Results and the $n=1$ Limitation}
\label{app:nslkdd}

This appendix reports the full BARS evaluation on
NSL-KDD~\cite{nslkdd2009}, separated from the main results
(Section~\ref{sec:results}) for methodological reasons explained
below. We summarize the dataset, the evaluation limitation that
motivates its appendix-only treatment, the per-method numbers, and a
brief reading of what the results do and do not show.

\subsection{Dataset and Protocol}

NSL-KDD is a refined version of the KDD~Cup~1999 dataset with
duplicate records removed and the train/test partitions held fixed
across studies. We use the standard 20\% training subset and the
canonical test split, with 41 features (numerical and one-hot encoded
categorical) and 20 classes (1 benign, 19 attack across the DoS,
Probe, R2L, and U2R families). After preprocessing
(Section~\ref{subsec:preprocessing}), the dataset contains 13{,}449
benign and 11{,}561 attack samples---a near-balanced regime
(1.16:1 benign-majority) and the smallest of the four benchmarks
considered in this work.

\subsection{Why NSL-KDD Is Reported Separately}

The standard NSL-KDD protocol provides a single predefined test set
and no canonical splitting strategy for cross-validation. Our
evaluation therefore produces \emph{one} test-set measurement per
(method, $k$) pair, in contrast to the five per-fold measurements
available on CICIDS2017 and CICDDoS2019. This has two consequences:

\begin{itemize}
    \item The paired Wilcoxon signed-rank test used in
    Section~\ref{subsec:significance} is uninformative on NSL-KDD:
    with $n=1$ paired observation, the test statistic is bounded such
    that no $p$-value below 0.5 is achievable regardless of effect
    magnitude. We omit NSL-KDD from Table~\ref{tab:wilcoxon} for
    this reason.

    \item Single-measurement comparisons are not robust to
    classifier-initialization variance. Apparent FPR differences of
    tens of percent between methods at adjacent feature budgets may
    reflect run-to-run noise rather than method differences, and
    cannot be distinguished without resampling.
\end{itemize}

We report NSL-KDD here for completeness and comparability with prior
NIDS feature-selection literature, but we do not draw method-level
conclusions from these results in the main paper.

\subsection{Per-Method Results}

Table~\ref{tab:nslkdd_main} reports FPR, TPR, and Macro-F1 for all
filter-based methods on NSL-KDD across the five feature budgets.

\begin{table}[t]
\caption{NSL-KDD: filter methods across feature budgets. Single
test-set measurement per cell. The wide budget-to-budget swings on
some methods illustrate the run-to-run variance that motivates
appendix-only reporting.}
\label{tab:nslkdd_main}
\centering
\small
\setlength{\tabcolsep}{4pt}
\begin{tabular}{ll ccccc}
\toprule
Metric & Method & $k{=}5$ & $k{=}10$ & $k{=}20$ & $k{=}30$ & $k{=}40$ \\
\midrule
\multirow{5}{*}{FPR$\downarrow$}
 & Pearson & 0.032 & 0.070 & 0.067 & 0.070 & 0.024 \\
 & MI      & 0.062 & 0.061 & 0.062 & 0.068 & 0.029 \\
 & Fisher  & 0.017 & 0.025 & 0.022 & 0.067 & 0.024 \\
 & CMD     & 0.064 & 0.046 & 0.070 & 0.070 & 0.025 \\
 & BARS    & 0.092 & 0.024 & 0.031 & 0.069 & 0.068 \\
\midrule
\multirow{5}{*}{TPR$\uparrow$}
 & Pearson & 0.670 & 0.702 & 0.736 & 0.709 & 0.702 \\
 & MI      & 0.650 & 0.650 & 0.701 & 0.697 & 0.706 \\
 & Fisher  & 0.678 & 0.699 & 0.712 & 0.701 & 0.707 \\
 & CMD     & 0.707 & 0.705 & 0.706 & 0.699 & 0.702 \\
 & BARS    & 0.755 & 0.696 & 0.699 & 0.697 & 0.710 \\
\midrule
\multirow{5}{*}{F1$\uparrow$}
 & Pearson & 0.154 & 0.211 & 0.273 & 0.332 & 0.361 \\
 & MI      & 0.175 & 0.261 & 0.264 & 0.300 & 0.343 \\
 & Fisher  & 0.239 & 0.303 & 0.376 & 0.338 & 0.351 \\
 & CMD     & 0.206 & 0.228 & 0.298 & 0.311 & 0.364 \\
 & BARS    & 0.157 & 0.213 & 0.316 & 0.312 & 0.344 \\
\bottomrule
\end{tabular}
\end{table}

\subsection{Reading the Results}

The BARS--CMD comparison on NSL-KDD exhibits wide and inconsistent
swings across budgets: BARS reduces FPR by $48.6\%$ at $k=10$ and
$55.3\%$ at $k=20$, but increases FPR by $45.1\%$ at $k=5$ and
$172.7\%$ at $k=40$. With only one measurement per cell, we cannot
attribute these swings to genuine method differences rather than to
run-to-run variance from the underlying MLP classifier's
initialization. The behavior of Fisher score and Pearson on the same
dataset shows similar instability (Fisher FPR ranging from 0.017 at
$k=5$ to 0.067 at $k=30$), reinforcing that the variance source is
the evaluation protocol rather than any particular method.

We note one observation that does emerge robustly: BARS preserves
TPR on NSL-KDD (range 0.696--0.755) at levels comparable to CMD
(0.699--0.707) and the other filters. The detection-capability claim
that holds on the three primary datasets also holds on NSL-KDD, even
though FPR comparisons cannot be made reliably at the single-run
level.

A more powered NSL-KDD evaluation---multi-seed runs over randomized
classifier initialization---is a clear next step. We chose not to
include such results in this paper to maintain a uniform evaluation
protocol across datasets (CICIDS2017 and CICDDoS2019 are evaluated
via per-fold variance, not per-seed), but it is a natural follow-up
for future work on benign-anchored feature selection.
\end{document}